\documentclass[11pt]{article}
\usepackage{moriond,epsfig,amssymb}

\def\be{\begin{equation}}
\def\ee{\end{equation}}
\def\bea{\begin{eqnarray}}
\def\eea{\end{eqnarray}}

\def\Dpgg{\Delta\phi_{\gamma\gamma}}
\begin{document}
\hfill {\tt  hep-ex/0606063} \\
\phantom{a} \hfill {\tt  FERMILAB-CONF-06-214-E}
\vspace*{2.4cm}
\title{PHOTON CROSS SECTIONS AT ECM = 2\,TEV}

\author{ M. WOBISCH%
~\footnote{Presented at:  XLIrst Rencontres de Moriond on ``QCD and 
 High Energy Hadronic Interactions'', La Thuile, Italy, March 18-25, 2006.}
 \\ (on behalf of the CDF and D\O\ collaborations) }

\address{Fermi National Accelerator Laboratory, 
Batavia, Illinois 60510, U.S.A.}

\maketitle\abstracts{        
Photon production rates have been studied
by the D\O\ and CDF experiments in Run II
of the Fermilab Tevatron Collider.
Measurements of the inclusive isolated
photon cross section and the di-photon
cross section are presented, 
based on integrated luminosities of
0.3\,fb$^{-1}$ and 0.2\,fb$^{-1}$, respectively.
The results are compared to perturbative
QCD calculations in various approximations.}

\section{Introduction} 

Photon cross sections in hadron collisions receive contributions
from ``prompt'' photons which directly emerge from the hard subprocess, 
and from photons which are produced during fragmentation.
The latter are usually accompanied by hadrons, and their contribution 
can be significantly reduced by requiring the photon to be isolated 
from other particles in the event.
Isolated photon cross sections are therefore dominated by prompt photons
and are directly sensitive to the dynamics of the hard subprocess and
to the strong coupling constant $\alpha_s$ and the parton density 
functions (PDFs) of the hadrons.
Furthermore, di-photon final states are also signatures for various 
``new'' physics processes, such as Large Extra Dimensions and for 
heavy new particles, such as the Higgs boson, decaying into photons.

Photon production has been considered an ideal source of direct 
information on the gluon density in 
the proton~\cite{Lai:1996mg,Martin:1996ev}.
However, it was observed~\cite{Aurenche:1998gv} that not all 
experimental data are consistently described by perturbative QCD (pQCD)
calculations in fixed order of the strong coupling constant, 
$\alpha_s$.
On one hand, it was argued that the existing data may not 
be consistent~\cite{Aurenche:1998gv}.
It was also suggested that the phenomenological introduction of 
intrinsic transverse momentum of the incoming partons may help 
improve the description of the data and reduce the 
inconsistencies~\cite{Martin:1998sq}.
But this ad-hoc procedure still had a significant model dependence
and was not seen to be a fundamental solution.
In recent global PDF fits photon data have been 
excluded~\cite{Martin:2004ir,Pumplin:2002vw}.

In order to rescue the photon data as a source of additional 
information on the PDFs, it is important either to identify
critical missing pieces in theory or to clearly establish 
inconsistencies of existing data sets.
Precision measurements in new kinematic regions are vital for 
testing theory predictions.
In this article, we present recent results from the D\O\ and
the CDF experiments on the inclusive isolated photon cross section
and the di-photon cross section, measured in proton-antiproton 
collisions at a center-of-mass energy of $\sqrt{s}=1.96\,$TeV 
during Run IIa of the Fermilab Tevatron Collider.

\section{Isolated Photon Cross Section}

\begin{figure}
\centering
\epsfig{figure=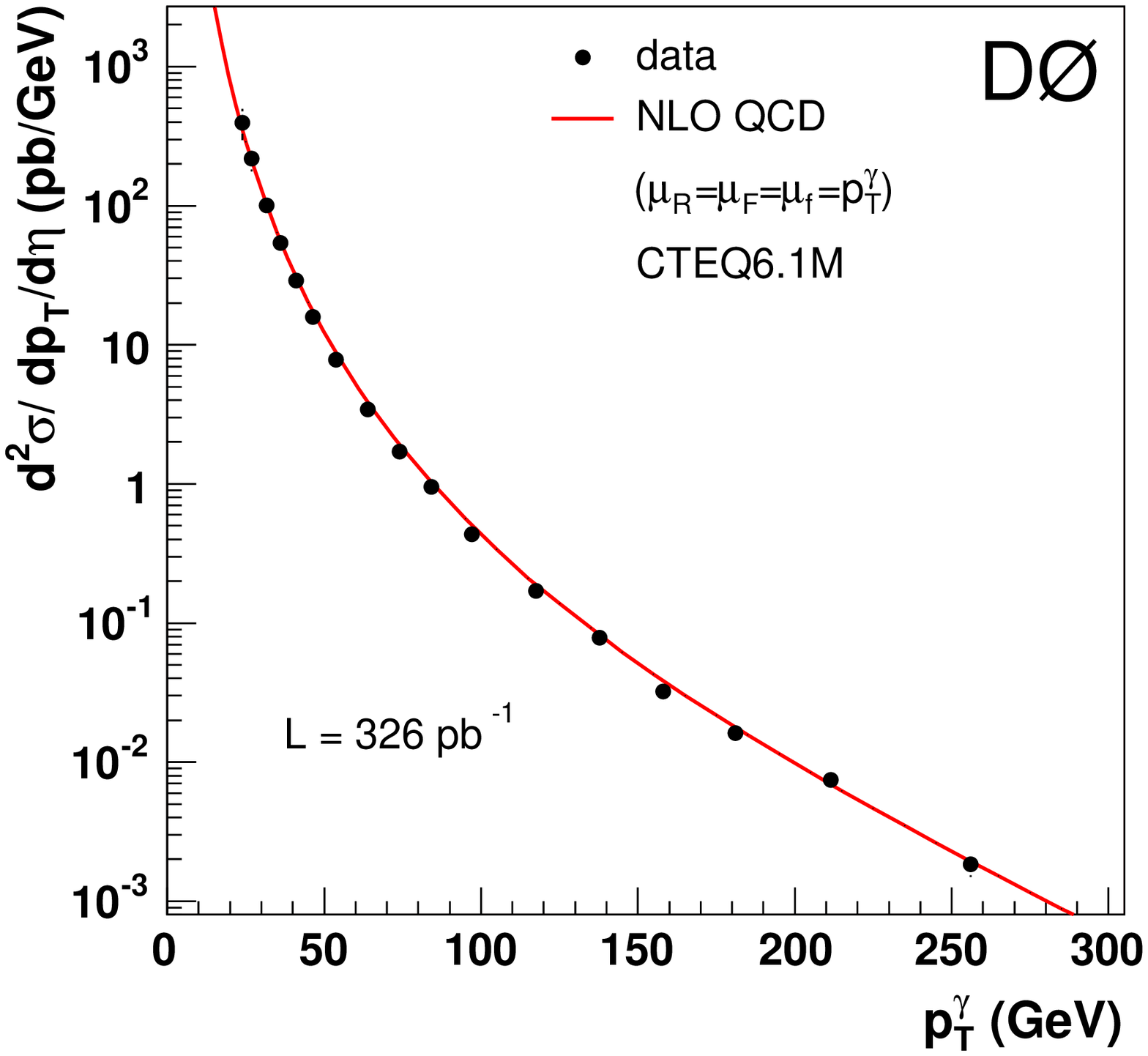,height=2.8in}
\epsfig{figure=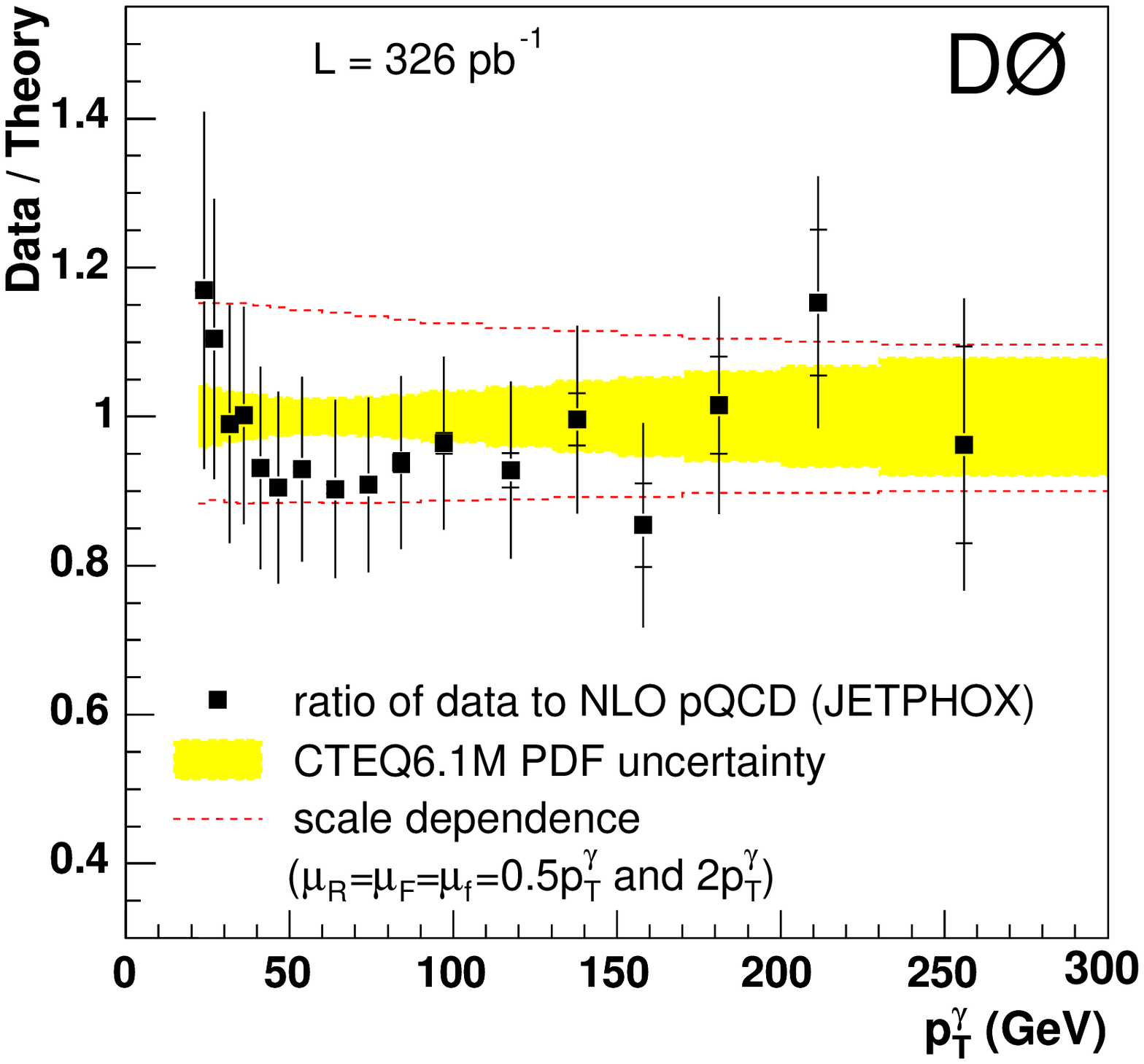,height=2.8in}
\caption{The isolated photon cross section as a function of the 
transverse photon momentum $p_T$ (left) and the ratio of data and
the NLO pQCD calculation from JETPHOX (right).
\label{fig:d0}}
\end{figure}

At lowest order, prompt photons are produced via quark-gluon Compton 
scattering or quark-antiquark annihilation.
Additional contributions to the inclusive photon cross section stem 
from the fragmentation of energetic $\pi^0$ and $\eta$ mesons.
In the D\O\ analysis~\cite{Abazov:2005wc}, photon candidates are 
defined as clusters of electromagnetic calorimeter cells
within a cone of radius $\Delta R=\sqrt{\Delta\eta^2 + \Delta\phi^2}<0.2$,
if more than 95\% of the energy is deposited in the electromagnetic 
part of the calorimeter and the probability of a matching track 
is below 0.1\%.
Cosmic backgrounds and electrons from $W$ decays are removed 
by requiring that the missing transverse energy in the event 
is less than 70\% of the transverse 
photon momentum ($p_T$).
Photons with $p_T > 23\,$GeV are selected in the range of 
pseudorapidities $|\eta|<0.9$.
The isolation criterion requires that the energy in a cone of 
radius $\Delta R <0.4$ not associated with the photon is less 
than 10\% of the photon energy.
The main background in the measurement are hadronic jets with a 
large electromagnetic fraction.
A neural net is trained to discriminate between photons and 
background using four sensitive variables~\cite{Abazov:2005wc}.
The photon purity was determined as a function of $p_T$ by fitting the 
neural net output for Monte Carlo signal and background to the observed 
distribution of photon candidates.
The photon $p_T$ distribution is corrected for energy resolution in an 
iterative unsmearing procedure.
The largest contributions to the experimental uncertainty are due to 
the purity determination and the photon energy calibration.

The results are based on 2.7 million photon candidates in an event 
sample corresponding to an integrated luminosity of 
$L_{\rm int}\simeq 0.3\,$fb$^{-1}$.
The isolated photon cross section is shown in Fig.~\ref{fig:d0} (left).
The cross section falls over five orders of magnitude in 
$23 < p_T < 300\,$GeV, the widest $p_T$ range ever covered by a 
single experiment.
The pQCD predictions are computed in next-to-leading order (NLO) 
in $\alpha_s$ using the program JETPHOX~\cite{Binoth:1999qq,Catani:2002ny}
with CTEQ6.1M PDFs~\cite{Pumplin:2002vw}, 
BFG fragmentation functions~\cite{Bourhis:1997yu}, and 
renormalization, factorization and fragmentation scales set to 
$\mu_{R,F,f}=p_T$.
The ratio of data and theory is shown in Fig.~\ref{fig:d0} (right),
and it is obvious that NLO theory gives a good description of the data
over the whole $p_T$ range.
The scale dependence of the NLO calculation is of the same size as 
the experimental uncertainties.
Uncertainties from the proton PDFs are significantly smaller over 
most of the $p_T$ range.
PDF sensitivity can therefore only be achieved if both experimental 
and theoretical uncertainties can be significantly reduced.

\section{Di-Photon Cross Section}

\begin{figure}
\centering
\epsfig{figure=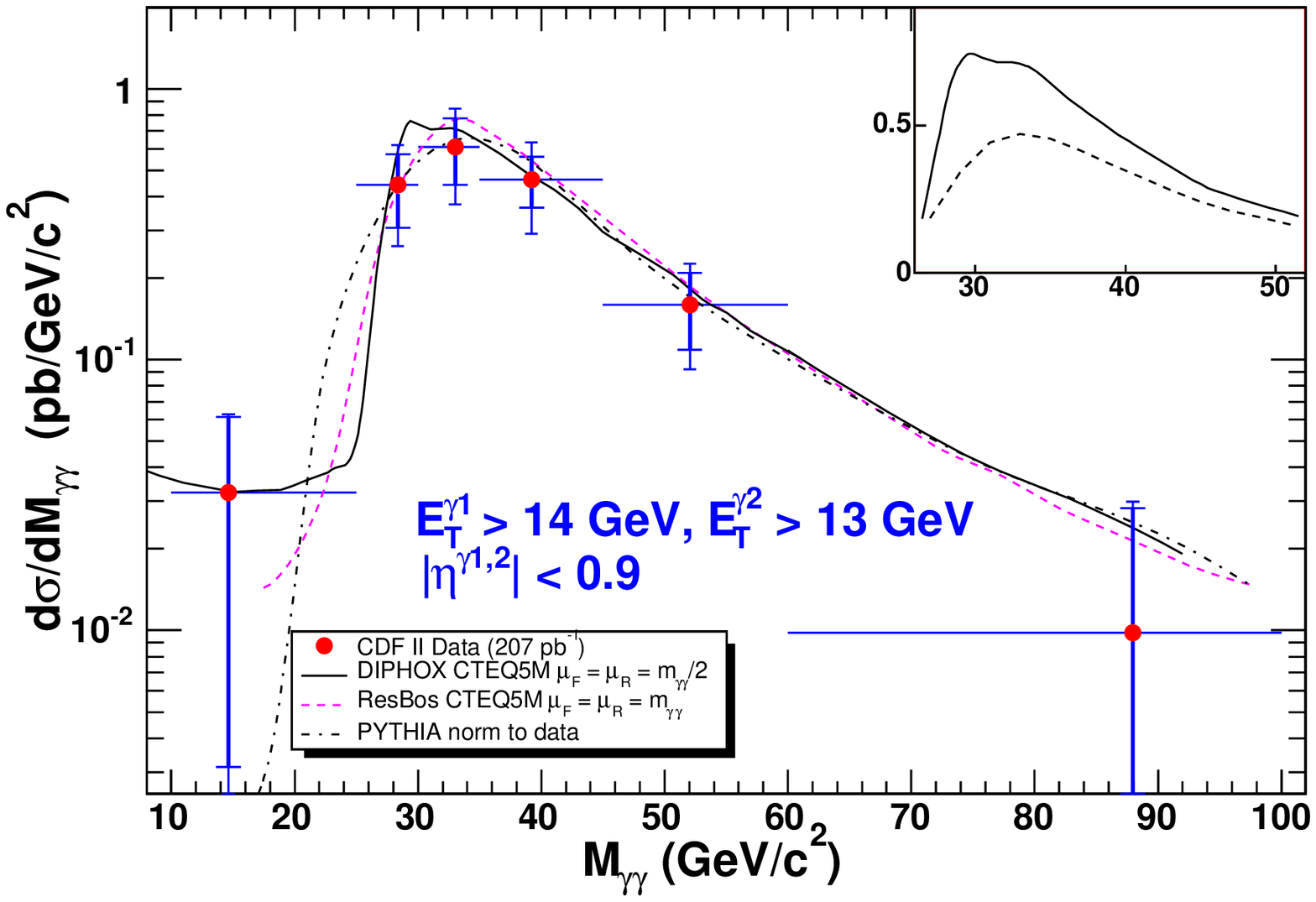,clip=,height=2.1in}\hskip-2mm
\epsfig{figure=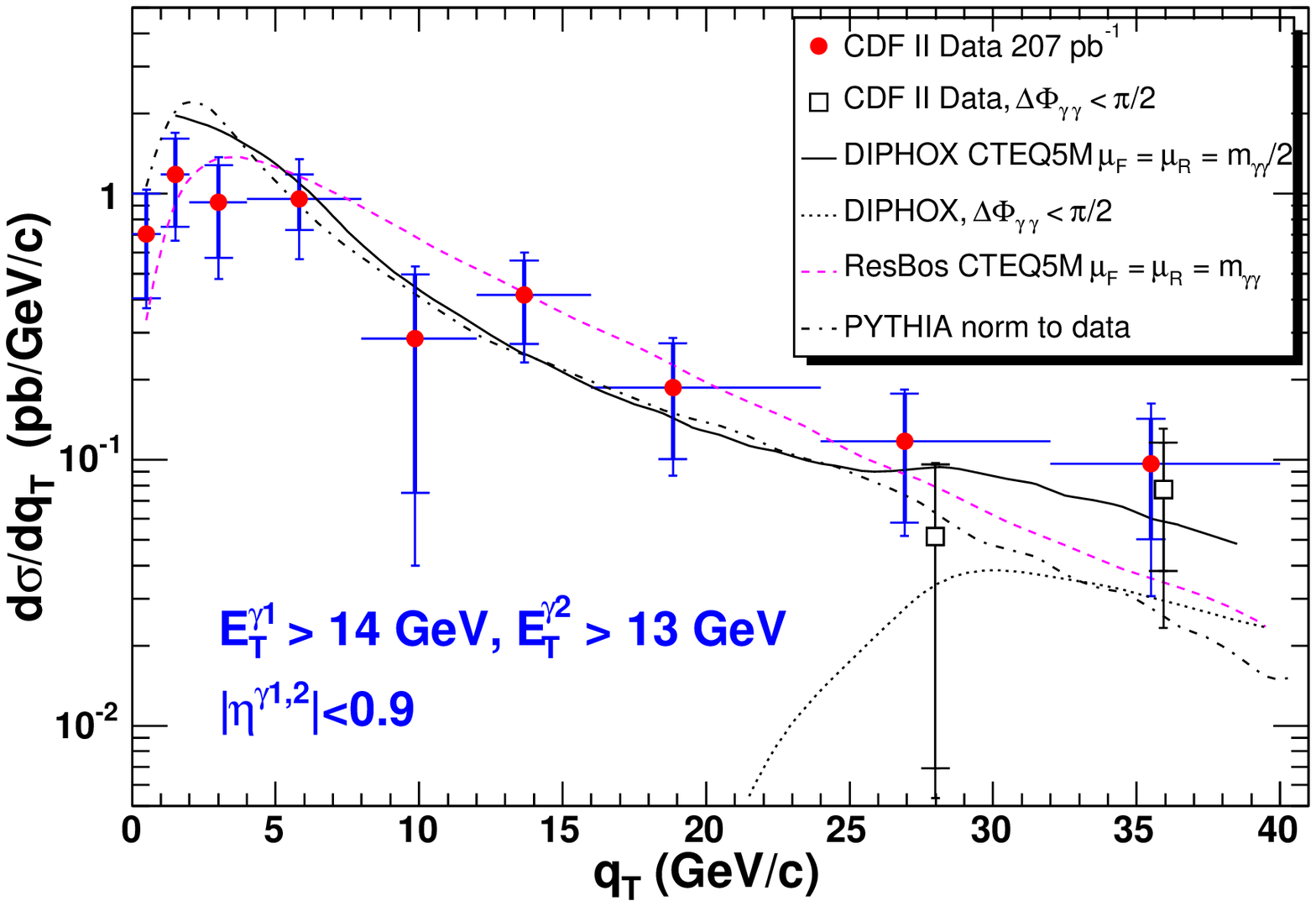,clip=,height=2.1in}
\epsfig{figure=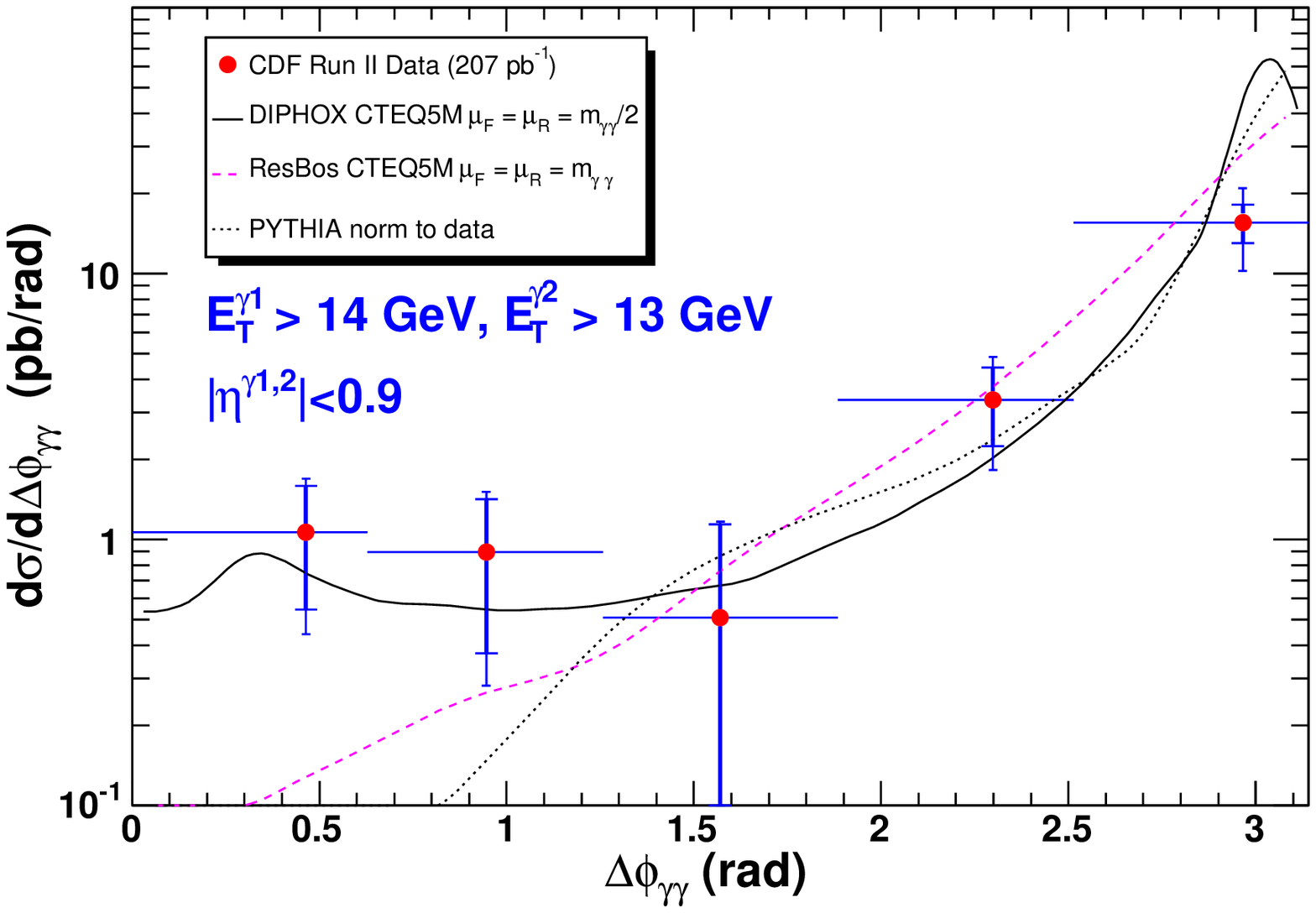,clip=,height=2.1in}
\caption{The CDF measurement of the di-photon cross section
as a function of the di-photon mass $M_{\gamma\gamma}$, 
the vector-sum of the transverse photon momenta $q_T$,
and the difference in azimuthal angle $\Dpgg$. 
\label{fig:cdf}}
\end{figure}

The leading contributions to di-photon production are from 
quark-antiquark annihilation and from gluon-gluon-scattering.
Although the latter process is suppressed by a factor of $\alpha_s^2$ 
(as the photons couple to a quark box), its contribution is still 
significant at small di-photon masses.
In this kinematic range the PDFs are probed at small momentum fractions,
where the gluon density is much larger than the quark densities.
In addition to the prompt contributions, the di-photon cross section 
also receives contributions where one or two photons are produced 
in fragmentation processes.
The CDF analysis~\cite{Acosta:2004sn} selects photon candidates based 
on the lateral shower profile and a veto against a matching track.
The isolation criterion requires the transverse energy in a cone radius 
of $R=0.4$ around the photon direction, not associated with the photon,
to be below 1\,GeV.
In total $427\pm59$ (stat.) di-photon events in 889 di-photon candidates 
are selected in the data sample corresponding to 
$L_{\rm int}\simeq 0.2$\,fb$^{-1}$.

The di-photon cross section is presented in Fig.~\ref{fig:cdf} as a 
function of di-photon mass (top left), the di-photon transverse momentum 
$q_T$ (top right), and the difference in azimuthal angle $\Dpgg$ between 
the two photons (bottom).
The results are compared to different approximations of pQCD.
The program DIPHOX~\cite{Binoth:1999qq} includes NLO matrix elements 
for both the direct contribution and the fragmentation contribution,
and the ${\cal O}(\alpha_s^3)$ corrections~\cite{Bern:2002pv}
to $gg\rightarrow\gamma \gamma$ have been added here to the 
calculation.
The program ResBos~\cite{Balazs:1997hv} has implemented 
the fragmentation contributions only at LO but it includes 
resummation of soft initial-state gluon radiation.
Also shown are the results for PYTHIA~\cite{pythia} 
(LO matrix elements plus parton shower and fragmentation model) 
which have been scaled by a factor of two.
All calculations give a reasonable overall description of the data, 
except in specific ``critical kinematic regions'' in which
the unique features of the different calculations are probed.
The contributions from fragmentation processes are especially large
in the regions of small di-photon masses, large $q_T$, and small $\Dpgg$.
These regions are only described by DIPHOX which includes
the NLO corrections for the fragmentation process.
The $q_T$ and the $\Dpgg$ distributions are directly sensitive 
to higher-order effects, since in lowest order both are trivial
($q_T=0$, $\Dpgg=\pi$).
The kinematic regions at low $q_T$ and large $\Dpgg$ are especially 
sensitive to soft initial-state gluon emissions.
Only ResBos describes this phase space due to the resummation of 
initial-state gluon radiation effects.
PYTHIA is not only too low by a factor of two, but it also fails in 
all ``critical regions''.
For an overall description of di-photon production a full NLO calculation 
(for direct and fragmentation contributions), combined with 
${\cal O}(\alpha_s^3)$ $gg\rightarrow \gamma \gamma$
corrections plus resummed initial-state contributions would 
be required.

\section{Summary}

The CDF and D\O\ experiments have published the first measurements of 
the single inclusive photon cross section and the di-photon cross section
at a center-of-mass energy of $\sqrt{s}=1.96\,$TeV.
The new inclusive isolated photon data extend previous measurements 
and indicate that the $\sqrt{s}$ and the $p_T$ dependence 
of direct photon production can be adequately described 
in NLO accuracy~\cite{Aurenche:2006vj}.
In the future, PDF sensitivity can only be achieved, if both
experimental and theoretical uncertainties can be significantly 
reduced.
The di-photon mass, $q_T$ and $\Dpgg$ dependence of the di-photon
cross section can not be simultaneously described by a single existing
theory calculation.
In different regions of phase space, different contributions are required: 
full NLO accuracy, in the direct and the fragmentation contributions,
resummed initial-state contributions and the ${\cal O}(\alpha_s^3)$
corrections to $gg\rightarrow\gamma \gamma$.
Since all the single ingredients are available, it should not be 
difficult to combine these into a powerful di-photon prediction.

\end{document}